%% file: conference_101719.tex
\documentclass[conference]{IEEEtran}
\IEEEoverridecommandlockouts
\usepackage{cite}
\usepackage{amsmath,amssymb,amsfonts}
\usepackage{algorithmic}
\usepackage{graphicx}
\usepackage{textcomp}
\usepackage{xcolor}
\usepackage{color}
\usepackage{graphicx}
\usepackage{graphics}
\usepackage{textcomp}
\usepackage{todonotes}
\usepackage{url}
\usepackage{float}
\usepackage{cleveref}
\usepackage[acronyms]{glossaries}

\newacronym{eutran}{E-UTRAN}{Evolved UMTS Terrestrial Radio Access Network}
\newacronym{ue}{UE}{User Equipment}
\newacronym{enodeb}{eNodeB}{Evolved NodeB}
\newacronym{epc}{EPC}{Evolved Packet Core}
\newacronym{ecgi}{ECGI}{E-UTRAN Cell Global Identifier}
\newacronym{mno}{MNO}{Mobile Network Operator}
\newacronym{rrc}{RRC}{Radio Resource Control}
\newacronym{mib}{MIB}{Master Information Block}
\newacronym{sib}{SIB}{System Information Block}
\newacronym{prach}{PRACH}{Physical Random Access Channel}
\newacronym{ie}{IE}{Information Element}
\newacronym{rach}{RACH}{Random Access CHannel}
\newacronym{ig}{IG}{Information Group}
\newacronym{ran}{RAN}{Radio Access Network}
\newacronym{ric}{RIC}{RAN Interface Controller}
\newacronym{qoe}{QoE}{Quality of Experience}
\newacronym{iot}{IoT}{Internet of Things}
\newacronym{rf}{RF}{Radio Frequency}
\def\BibTeX{{\rm B\kern-.05em{\sc i\kern-.025em b}\kern-.08em
    T\kern-.1667em\lower.7ex\hbox{E}\kern-.125emX}}
\begin{document}

\title{Scrutinizing Real-life Configurations of Random Access Procedures in Cellular Networks\\
\thanks{This research was supported by the National Growth Fund through the Dutch 6G flagship project “Future Network Services”.}
}%

\author{\IEEEauthorblockN{Joris Belder}
\IEEEauthorblockA{\textit{D-INFK} \\
\textit{ETH}\\
Zurich, Switzerland \\
jbelder@student.ethz.ch}
\and
\IEEEauthorblockN{Anup Bhattacharjee}
\IEEEauthorblockA{\textit{Networked Systems} \\
\textit{Delft University of Technology}\\
Delft, the Netherlands \\
a.k.bhattacharjee@tudelft.nl}
\and
\IEEEauthorblockN{Fernando Kuipers}
\IEEEauthorblockA{\textit{Networked Systems} \\
\textit{Delft University of Technology}\\
Delft, the Netherlands \\
f.a.kuipers@tudelft.nl}
}

\maketitle

\begin{abstract}
In cellular networks, base stations broadcast configurations that devices use for the random access procedure, which is a vital part of the connection setup.
Ideally, the network should choose configurations based on the deployment scenario to optimize radio resource management.
Doing so can, for example, decrease collisions of random access messages.
We captured 112,806 data points of cellular broadcast information from nine network operators across three countries
and analyzed how the operators configure the random access procedure.
We found that configurations often do not fit the deployment scenario, and neighboring cells often use the same configuration,
causing an unnecessarily high risk of collisions and, hence, delay in the connection setup.
Furthermore, we simulated the random access procedure in NS-3 and found that by varying the configurations in a large area with many cells, the number of collisions can be reduced by 43\% on average and up to 61\%, and the connection delay can be lowered by 11\% on average and up to 42\%.
Our findings indicate that simple adaptations in the random access configurations can greatly improve the performance of cellular networks.
\end{abstract}


\maketitle

\input{sections/introduction}
\input{sections/background}
\input{sections/measurements}
\input{sections/simulations}
\input{sections/relatedWork}

\section{Discussion}
Our measurements were obtained from 4G networks and are directly relevant to today’s cellular infrastructure: in 2024 about 58\% of smartphone subscribers worldwide still used 4G~\cite{gsma2025}, and in Europe the vast majority of 5G deployments are non-standalone (NSA) relying on LTE for control. Indeed, only 2\% of European Speedtest samples are on 5G standalone (SA) networks~\cite{ookla2025}, underscoring the continued dominance of 4G and the need to optimize its operation. 

Because mobile operators do not make data on connection collisions or delays available from live networks, we could not evaluate the effect of our PRACH reconfiguration scheme in a live network. Instead, like many other works, we resorted to packet-level simulations in NS‑3: we implemented the 3GPP random-access procedure using the NS-3.24 simulator with the LENA+ PRACH module. This approach yields a much more realistic view of network behaviour than a purely analytical model: NS‑3/LENA+ explicitly models the PHY/MAC stack (including path-loss, fading, and collision effects) and simulates user arrivals (e.g., a uniform population of UEs attempting simultaneous access). In contrast, closed-form predictions typically omit these details. Of course, even NS‑3 has its limitations. For example, LENA+ only supports a subset of PRACH‐ConfigIndex values (even-frame transmissions and format 0) and is tied to an older NS-3 release. These constraints highlight the need for richer simulation tools; in particular, we encourage a community effort to extend NS‑3 (or ``LENA'') to support 5G-Standalone features so that next-generation RAN behaviour can also be accurately evaluated.

\section{Conclusion}
In this paper, we have measured, analyzed, and simulated the random access configurations of real-life cellular networks. Our measurements show that random access configurations in the wild barely adapt to the deployment scenarios at hand, and neighboring cells often use the same configurations.
Our simulations reveal that such configurations are deficient, and assigning configurations more astutely can greatly reduce the risk of collisions, shortening the connection setup for many users, and allowing for more efficient use of network resources.
We hope our work encourages operators and researchers to design and deploy such astute random access configuration schemes,
and we envision the advent of \gls{ric}s to enable these schemes in practice.
Furthermore, we invite the community to explore the configurations of other cell broadcast parameters, for which there might be room for optimizations similar to what we have shown.

\section*{Acknowledgements}
This research was supported by the National Growth Fund through the Dutch 6G flagship project “Future Network Services”. Furthermore, we thank Adrian Zapletal for his valuable feedback on this work.

\bibliographystyle{IEEEtran}
\bibliography{refs}

\appendix

\section*{Ethics}
\label{ethics}
The 4G dongle \gls{ue} we used triggers and sends a failed attach procedure to the \gls{enodeb}, but it does not generate any traffic load and neither harms the mobile network infrastructure nor other users.
Additionally, we ensured that no authorized or unauthorized connections to any mobile networks were made during our experiments.

\section*{Information Elements for Random Access}
\label{appendix-ie-random-access}
\begin{figure}
    \centering
    \includegraphics[width=0.999\columnwidth]{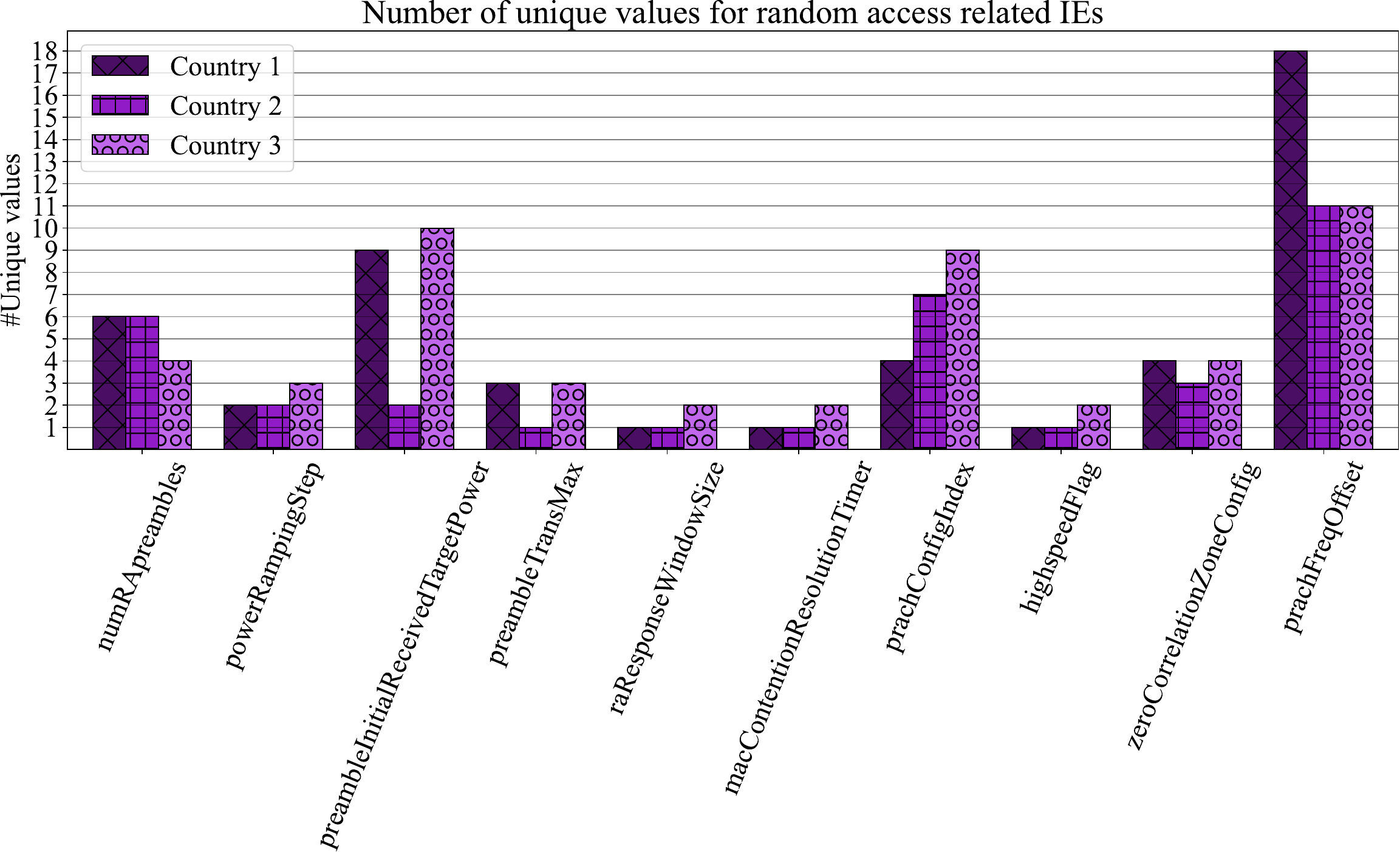}
    \caption{Number of unique values for different IEs in SIB2 pertaining to the random access procedure.}
    \label{fig:random-access-ies}
\end{figure}

As mentioned in \Cref{sec:measurements:insights}, SIB2 contains many \glspl{ie} related to random access for which we observed few different configurations.
\Cref{fig:random-access-ies} shows for the \glspl{ie} in SIB2 that pertain to random access how many unique values we observed in our measurements.
For approximately half of the \glspl{ie} we present here, we observed very few different values being used even though deploying different values based on the deployment scenario might be sensible.
Hereafter, we provide brief descriptions of each of these \glspl{ie}.

\noindent \textbf{numRApreambles}: Number of non-dedicated random access preambles~\cite{ts36-321}.

\noindent \textbf{powerRampingStep}: Power ramping factor. Value in dB~\cite{ts36-321}.

\noindent \textbf{preambleInitialReceivedTargetPower}: Initial preamble power. Value in dBm~\cite{ts36-321}.

\noindent \textbf{preambleTransMax}: Maximum number of preamble transmissions~\cite{ts36-321}. 

\noindent \textbf{raResponseWindowSize}: Duration of the random access response window. Value in subframes~\cite{ts36-321}.

\noindent \textbf{macContentionResolutionTimer}: Timer for contention resolution. Value in subframes~\cite{ts36-321}.

\noindent \textbf{prachConfigIndex}: The \gls{prach} configuration index that defines the preamble~\cite[\S5.7.1]{ts36-211}.

\noindent \textbf{highspeedFlag}: High-speed-flag. TRUE corresponds to Restricted set and FALSE to Unrestricted set~\cite[\S5.7.2]{ts36-211}.

\noindent \textbf{zeroCorrelationZoneConfig}: $N_{CS}$ configuration, which is the cyclic shift value used for random access preamble generation~\cite[\S5.7.2]{ts36-211}.

\noindent \textbf{prachFreqOffset}: The \gls{prach} frequency offset~\cite[\S5.7.1]{ts36-211}.

\end{document}

%% file: sections/introduction.tex
\section{Introduction}
\label{sec:intro}
In 2024, about 58\% of global smartphone subscribers still used 4G~\cite{gsma2025}, excluding those on 5G non-stationalone (NSA) networks that rely on 4G for the control plane. In contrast, adoption of 5G Standalone (SA), where both user and control planes operate on 5G, remains limited in Europe. According to Speedtest, only 2\% of samples in Europe were on 5G SA networks, significantly behind China (80\%), India (52\%), and the United States (24\%)~\cite{ookla2025}. This shows the high relevance of 4G networks and the use of control methods to optimize the configurations of the cellular networks.

One vital aspect of cellular networks is \emph{connection setup}.
Devices perform a \emph{random access procedure} to initialize a connection and receive uplink transmission grants.
While researchers have already presented novel random access schemes~\cite{lenaplus,huang2017r}, it remains unclear how actual cellular deployments configure the random access procedure.
Such insights into real-life configurations are essential for understanding and improving the performance of cellular networks and, in particular, of the connection setup.

In this paper, we present measurements of random access configurations in 4G networks.
Our measurements encompass 112,806 data points and stretch across three countries.
In each country, we took measurements for the three largest \glspl{mno} by subscriber base.
Specifically, we collected cell broadcast information that devices use to configure the random access procedure.
For instance, cells broadcast information that tells devices which preamble they should use for initializing connections.
To the best of our knowledge, we are the first to present such measurements.

Our data indicates that \emph{random access configurations often do not suit the deployment scenario and are often the same across many cells}.
For example, neighboring cells commonly use the same preamble configuration for random access, which \emph{unnecessarily increases the risk of collisions, prolongs the connection setup, and leads to inefficient network resource usage}.
To analyze how much of an issue cells using the same configuration
is, we simulated the random access procedure in NS-3 and compared the performance when cells use the same configuration with the performance when cells use different configurations.
We found that \emph{simply deploying different configurations across cells reduces the number of collisions and lowers the connection delay significantly}.
For example, one can lower the median connection delay by up to 300\,ms by deploying simple changes.
\glspl{mno} can leverage our findings to improve the performance of cellular connection setup and to save network resources.

While we measured the configurations in 4G networks, the random access procedure is similar in 5G. 
When considering Europe, most 5G deployments are currently non-standalone, meaning they have 4G as their control layer~\cite{ookla2025}.
Our insights therefore also hold for more modern networks.
As pushed by the Open-\gls{ran} initiatives~\cite{oran}, future networks will be more softwarized, with the \gls{ran} being controlled by a \gls{ric}~\cite{oran-near-rt-ric}.
We envision \glspl{ric} to adjust the configurations of cells intelligently to improve network performance, and part of these configurations is the random access procedure, which we show can be a crucial factor in the performance of cellular networks. O-RAN provides interfaces to include 4G base stations as part of the O-RAN architecture, which will allow to make adjustments to the configurations for improved performance.

\noindent\textbf{Contributions.}\quad
Our contributions are as follows:
\begin{itemize}
    \item We present data on real-life random access configurations and show that configurations often do not suit the deployment scenario and are often the same across cells.
    \item We use NS-3 simulations to show that using different configurations across cells can notably reduce the number of collisions and the connection delay.
\end{itemize}

%% file: sections/background.tex
\section{Background}
\label{sec:background}
\noindent\textbf{Basics.}\quad
In 4G, \glspl{ue}, e.g., mobile phones, connect to base stations called \glspl{enodeb}, which connect to an \gls{epc}. 
The \gls{epc} is 
responsible for services such as user authentication, connection setup, and roaming~\cite{ts23-002}.
Each \gls{mno} has their own frequency bands~\cite{gsma:spectrum-regional} to transmit and receive signals. This means that \glspl{mno} are not allowed not transmit in each others spectrum.
To efficiently utilize these bands, \glspl{mno} deploy \glspl{enodeb}. Each \gls{enodeb} divides the allocated frequency bands into multiple logical cells. These logical cells operate on different frequencies to minimize interference within the same \gls{mno} network.

A \gls{ue} connects to one of these logical cells by communicating over a designated frequency band during specific time intervals known as radio frames. From this point onward in the paper, we refer to the ``logical cell'' as a ``cell''.

\par In 4G, the duration of a radio frame is 10\,ms.
Each radio frame is further divided into ten subframes of equal length.
Signals sent on the same frequency and in the same subframe can collide~\cite{ts36-211}.
Therefore, \glspl{mno} should optimize their cellular configurations to prevent collisions.

\begin{figure}
    \centering
    \includegraphics[width=\columnwidth]{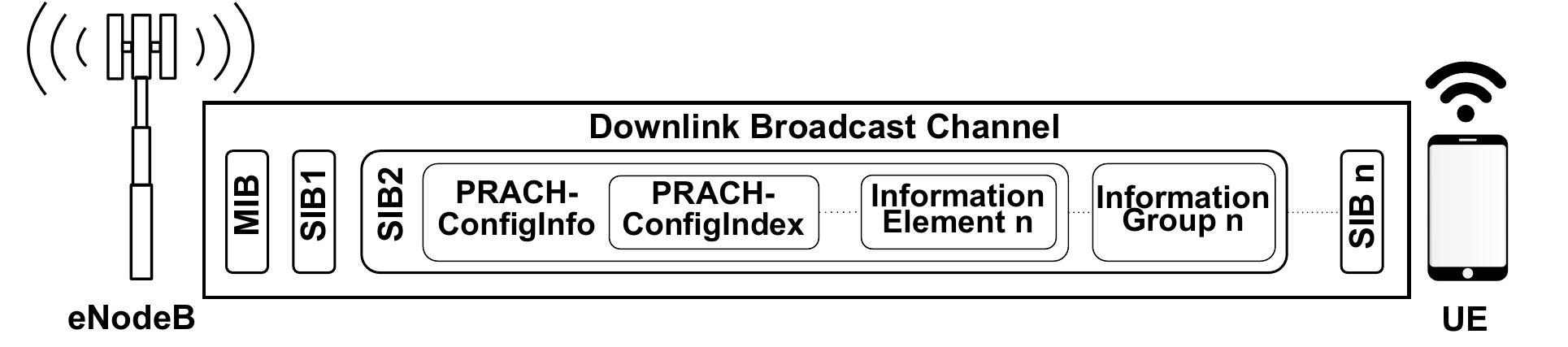}
    \caption{Transmission of the Master and System Information Blocks.
    SIB2 contains information on how to perform the random access procedure.}
    \label{fig:sib-structure}
\end{figure}

\noindent\textbf{Cell Broadcast.}\quad
Cells broadcast system information using the \gls{rrc} protocol~\cite{ts36-300},
which manages network resources pertaining to user connections.
With \gls{rrc}, cells regularly broadcast a \gls{mib} that contains information that \glspl{ue} need to synchronize and listen to other broadcast information such as different types of \glspl{sib}.
\glspl{sib} provide further information for \glspl{ue} to access cells.
The \gls{mib} and the \glspl{sib} are divided into multiple \glspl{ig} that are further subdivided into \glspl{ie}.
Each \gls{ie} typically describes some \gls{ran} parameter and assists the \glspl{ue} in choosing configuration settings~\cite{ts36-331}.
SIB1 contains information on the scheduling of other \glspl{sib}.
SIB2 contains radio resource configuration information
that \glspl{ue} use for several procedures, including random access.
\Cref{fig:sib-structure} illustrates the transmission of the \gls{mib} and \glspl{sib} and parts of the structure of SIB2.

\noindent\textbf{Random Access Procedure.}\quad
To connect or reconnect to a cell and receive uplink grants for data transfer, \gls{ue}s perform a random access procedure.
They must send a \gls{prach} preamble to the cell to establish the connection.
To decide which preamble to use, \gls{ue}s listen to the cell broadcast information.
SIB2 contains the \gls{ig} \textit{PRACH-ConfigInfo}, and this \gls{ig} contains \gls{ie} \textit{PRACH-ConfigIndex} (as illustrated in \Cref{fig:sib-structure}) that tells \glspl{ue} which preamble to use and when to send it.
There exist four \gls{prach} preamble formats that suit different deployment scenarios.
For instance, preamble formats 0 and 2 are suitable for smaller cells (such as those in urban deployments), whereas formats 1 and 3 suit larger cells (such as those in rural deployments), because the latter use longer guard intervals, giving signals more room to travel~\cite{ts36-211,lte_for_umts,self_organising_networks}.
For each \gls{prach} preamble format, there are 16 different \gls{prach} preambles.
Preambles differ in during which subframe numbers \glspl{ue} may transmit them and in whether \glspl{ue} may transmit them during odd, even, or all radio frame numbers.

If several \glspl{ue} attempt to connect at the same time with the same \gls{prach} preamble, the preambles may collide even if they were meant for different cells, causing one or multiple of the connection attempts to fail.
When a collision occurs, \glspl{ue} must retry connecting later.
So, collisions prolong the connection setup.
Thus, \glspl{mno} should configure their cells such that neighboring cells use different preambles to prevent collisions, and they should ensure that they use preamble formats that suit the deployment scenario.

%% file: sections/measurements.tex
\section{Measurements and Insights}
\label{sec:measurements}
We captured \gls{sib} messages from cells in three different countries and for three \glspl{mno} per country.
\Cref{sec:measurements:data-collection} details our setup for the data collection.
We analyzed the collected data to see how \glspl{mno} configure the random access procedure.
We detail our insights in \Cref{sec:measurements:insights}.

\subsection{Data Collection}
\label{sec:measurements:data-collection}
\begin{table}[]
\caption{Measurement Statistics}
\label{tab:stats}
\resizebox{\columnwidth}{!}{%
\begin{tabular}{lccc}
\hline
 &
  \textbf{Country 1} &
  \textbf{Country 2} &
  \textbf{Country 3} \\ \hline
MNOs &
  3 &
  3 &
  3 \\ \hline
Frequency bands observed &
  \begin{tabular}[c]{@{}c@{}}1, 3, 7, 8,\\ 20, 28, 38\end{tabular} &
  \begin{tabular}[c]{@{}c@{}}1, 3, 7, 8, \\ 20, 28, 38\end{tabular} &
  \begin{tabular}[c]{@{}c@{}}1, 3, 7, \\ 20, 28\end{tabular} \\ \hline
Municipalities measured &
  7 &
  7 &
  10 \\ \hline
\begin{tabular}[c]{@{}l@{}}eNodeBs measured\\ (grouped by \gls{mno} )\end{tabular} &
  155 &
  188 &
  43 \\ \hline
\begin{tabular}[c]{@{}l@{}}(Logical) Cells measured\\ (grouped by \gls{mno})\end{tabular} &
  992 &
  2036 &
  807 \\ \hline
Measurement locations &
  29 &
  62 &
  13 \\ \hline
Total measurements &
  23,389 &
  76,556 &
  12,861 \\ \hline
Measurement period &
  December 2022 &
  \begin{tabular}[c]{@{}c@{}}August - December \\ 2022\end{tabular} &
  December 2022 \\ \hline
\end{tabular}%
}
\end{table}

For our data collection, we used a laptop running Ubuntu 18.04 with Linux kernel 5.4.0-132-lowlatency
alongside a Quectel EG25-G USB dongle%
\footnote{We used firmware version \texttt{EG25GGBR07A07M2G}.}
as \gls{ue} that connects to cells.
Using this setup, we captured \gls{sib}1 and \gls{sib}2 broadcast messages from cells.
We developed the code for capturing these messages in Python 3.6.9 using QCSuper~\cite{qcsuper} at commit \texttt{5c4e529} and Tshark 2.6.10.
To ensure comprehensive coverage, we made use of the websites Cellmapper~\cite{cellmapper} and OpenCellid~\cite{opencellid},
which provide public information on cells such as their location, operator, and frequencies.
We used the websites to make a list of physical measurement locations where we could capture broadcast messages from multiple operators. 
Our measurements encompass a total of 112,806 data points.
We obtained most measurements from Country 2, which has the largest population, mobile user base, and number of cells among the three countries.
\Cref{tab:stats} presents detailed statistics about our measurements.

\subsection{Insights}
\label{sec:measurements:insights}

As explained in \Cref{sec:background}, \gls{prach} preambles meant for different cells can collide when they are sent at the same frequency (which depends on the \gls{mno} and the frequency band) and during the same subframe (which is determined by the \textit{PRACH-ConfigIndex}).
In what follows, we analyze our collected data and unveil how real-life configurations of cellular deployments incur an unnecessarily high risk of collisions.

\begin{figure}
    \centering
    \includegraphics[width=\columnwidth]{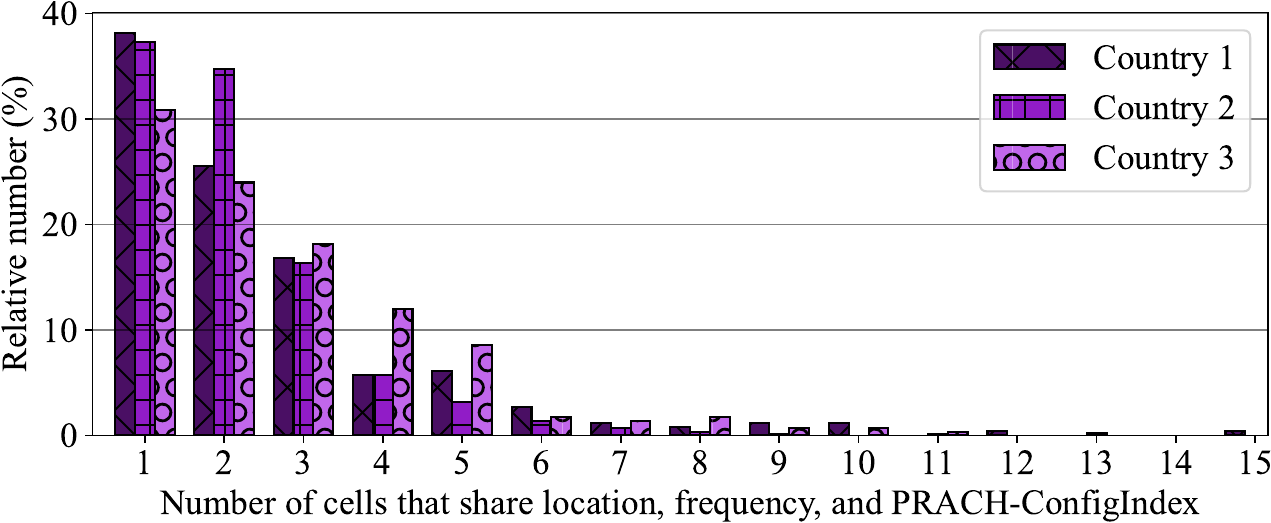}
    \caption{The relative number of cells that share location, frequency, and \textit{PRACH-ConfigIndex}, for all three countries.
    Often, several cells share these properties and can, thus, incur collisions during random access.}
    \label{fig:cells-per-grouping}
\end{figure}

\noindent\textbf{Insight 1: In many locations, there are multiple cells that share the same frequency and \textit{PRACH-ConfigIndex}, making collisions during random access possible.}\quad
For each measurement location, we determined the number of cells that share the same frequency and \textit{PRACH-ConfigIndex}.
\Cref{fig:cells-per-grouping} illustrates how often each number occurred per country, relative to the total number of measurements per country.
In all three countries, more than 60\% of the locations contain multiple cells that share the same frequency and \textit{PRACH-ConfigIndex} and are, therefore, prone to collisions during the random access procedure.
Indeed, this may be because of frequency reuse. But for different cells, a different value for PRACH-ConfigIndex can be used to reduce the probability of collision even in the case of frequency reuse.
The probability of collisions increases with the number of cells that share frequency and \textit{PRACH-ConfigIndex}.
In our data, 25\%--35\% of the locations contain three to five such cells, and the number of cells can reach up to fifteen.
Our findings indicate suboptimal configurations of cellular deployments;
\glspl{mno} could easily deploy different \textit{PRACH-ConfigIndex} values to reduce the risk of collisions and lower the connection delay.

\noindent\textbf{Insight 2: Many cells deploy the same \gls{prach} preamble format, which is not always suitable for the deployment scenario.}\quad
We found that the \glspl{mno} for which we collected data use only two out of four preamble formats: 0 and 1.
Format 0 supports cells with a range of up to $\sim$14\,km, whereas format 1 supports up to $\sim$75\,km due to its higher tolerance in fading environments~\cite{book:lte-umts} and its longer guard interval~\cite{book:lte-advanced} compared to format 0.
Hence, format 1 is better suited to larger cells and areas with less fading, such as cells in rural environments~\cite{tr38-913}.
To verify whether \glspl{mno} choose preamble formats based on the cell location, 
we classified our measurement locations into urban, suburban, and rural.
We call a location urban if it is within 100\,m of a city center, a busy public space (parks, tourist attractions, train stations, hospitals, or university buildings), or a main road that leads through a major city.
We call a location suburban if it is in a city but does not fall into our urban categorization.
We call a location rural if it is outside of city borders.
Our categorization is by no means a proper classification one could use for urban planning, but it suffices for our purpose.

\begin{figure}
    \centering
    \includegraphics[width=.9\columnwidth]{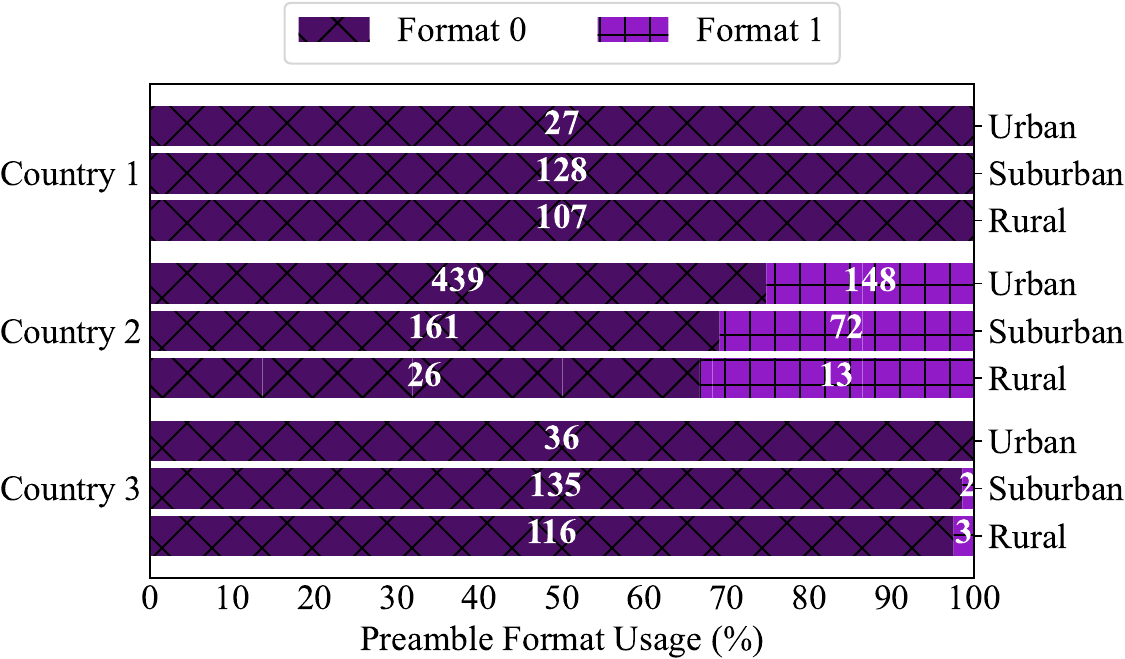}
    \caption{The usage of each preamble format of cells in urban, suburban, and rural areas, with the relative amount of cells using a preamble format on the x-axis and the absolute amount on the bars.
    Only two formats are in use, and the vast majority of cells use format 0.}
    \label{fig:format-0-usage}
\end{figure}

\Cref{fig:format-0-usage} presents how often we observed each preamble format in each type of location (urban, suburban, rural) per country.
The \glspl{mno} we measured predominantly deploy format 0.
In two countries, deployments exclusively or almost exclusively use format 0.
Country 2 sometimes uses format 1, but format 0 still dominates,
and there is little difference between rural and urban locations in the usage of format 1 even though the formats suit different types of locations better.
We found this trend to appear for all \glspl{mno}.

\noindent\textbf{Insight 3: For many \glspl{ie}, only one or few different configurations are deployed.}\quad
Our data shows that \glspl{mno} deploy few different configurations not only for the \textit{PRACH-ConfigIndex} but also for other \glspl{ie} in SIB2 that pertain to random access.
For example, for the \gls{ie} random access response window size (ra-ResponseWindowSize), which indicates how long \glspl{ue} should wait for a response after they sent the preamble, we observed but one or two unique values per country.
It would be sensible to adapt this value to the deployment scenario and, for instance, set a longer window for denser deployments.
For many other \glspl{ie}, we observed similarly few different values.
We present the details in \Cref{appendix-ie-random-access}.
While our work focuses on the \textit{PRACH-ConfigIndex}, which plays an important role in the random access procedure, these findings indicate that the configurations of cellular deployments are deficient in general.

%% file: sections/simulations.tex
\section{Simulations}
\label{sec:simulations}
In this section, we verify that the current assignment of the \textit{PRACH-ConfigIndex} IEs in the wild is deficient,
and we show how much the network performance can be improved with more thoughtful assignments.
To this end, we simulate the random access procedure with varying numbers of cells using NS-3~\cite{ns3} and evaluate the performance when cells use the same \textit{PRACH-ConfigIndex} and when they use different indices.
We use two metrics: the \emph{number of collisions}
(how often cells do not receive a \gls{prach} preamble due to interference with another preamble)
and the \emph{connection delay}
(the time it takes a \gls{ue} to finish the connection setup after sending the first \gls{prach} preamble).
To measure the potential improvement that using different indices can yield, we need data on the number of collisions and the connection delay experienced by \glspl{ue}.
However, \glspl{mno} do not make such data available from their deployments.
So, we turn to simulations to get an idea of what could be possible.

\subsection{Simulation Setup}
We implemented our simulations in NS-3.24~\cite{ns3.24} and used the \textit{LENA+} module~\cite{lenaplus}, which implements a \gls{prach} model in NS-3%
\footnote{\textit{LENA+} is built on top of NS-3.24 and supports only this version of NS-3, which is why we could not use a newer version of the simulator.}.
\textit{LENA+} adds support for the \textit{PRACH-ConfigIndex} \gls{ie} to NS-3 and turns the messages \glspl{ue} and cells exchange as part of the random access procedure into proper signals that are affected by path loss and collisions;
vanilla NS-3 does not properly model these signals.
Moreover, \textit{LENA+} allows us to collect the number of collisions and the connection delay as defined above.
A peculiarity of \textit{LENA+} is that it always sets the \textit{PRACH-ConfigIndex} \gls{ie} to 0.
Thus, we modified the module to allow for selecting different \textit{PRACH-ConfigIndex} values.

Our module can simulate a specific set of \textit{PRACH-ConfigIndex} values.
These values allow for preambles to be sent only during even radio frame numbers.
Furthermore, only preamble format 0 exists in \textit{LENA+}, effectively limiting the usable values for \textit{PRACH-ConfigIndex} to 0, 1, 2, and 15 because these values correspond to even radio frames and preamble format 0.
Adding support for other \textit{PRACH-ConfigIndex} values would require major changes to \textit{LENA+}'s scheduler, which was not needed because the usable values are precisely the ones we need, as we clarify hereafter.
In urban scenarios, many users are already connected to the network and are actively using it for data transmission.
Spending too many network resources (subframes and radio frames) on the random access procedure would come at the cost of already connected users.
Because the group of already connected users is typically larger, one should use \textit{PRACH-ConfigIndex} values that imply minimal time spent on the random access procedure to increase the throughput of data transmissions~\cite{lte_for_umts, self_organising_networks}.
The \textit{PRACH-ConfigIndex} values we can simulate are precisely the ones that imply minimal time spent on the random access procedure.
Hence, they do not limit our results.
Another point to keep in mind is that, in this study, our simulations do not consider frequency reuse. When a \gls{mno} deploys its network, it typically employs a frequency reuse pattern to minimize interference between cells. However, our focus for this paper has been on scenarios where a \gls{ue} receives \gls{sib} messages for the same \gls{mno} and frequency from different logical cells. As illustrated in Figure~\ref{fig:cells-per-grouping}, the number of logical cells sharing the same \textit{PRACH-ConfigIndex} can reach up to 15 in some cases. Investigating frequency reuse patterns could be a valuable avenue for future work or for other members of the research community to explore as an extension of this study.
Our simulation uses an urban path loss model provided by NS-3%
\footnote{Specifically, we used NS-3's HybridBuildingsPropagationLossModel.}
and contains only the setup that is necessary for the random access procedure.
So, we can realistically test this procedure for any number of \glspl{ue} and cells and for different configurations of the \textit{PRACH-ConfigIndex}.

\begin{table}
\centering
\caption{General simulation parameters.}
\resizebox{.55\columnwidth}{!}{%
\begin{tabular}{cc}
\hline
    \textbf{Parameter}        & \textbf{Value}                       \\ \hline
    Simulation repetitions    & 5                                    \\ \hline
    Simulation time           & 5\,s                                 \\ \hline
    UE height                 & 1\,m                                 \\ \hline
    Frequency band (downlink) & 17 (740\,MHz)                        \\ \hline
    Cell antenna height       & 30\,m                                \\ \hline
    Inter-cell distance       & 200\,m                               \\ \hline
\end{tabular}%
}
\label{tab:sim-params}
\end{table}

\Cref{tab:sim-params} shows the simulation parameters common to all simulations.
We set most parameters in accordance with the 3GPP specification 
on 4G deployment scenarios~\cite{tr38-913}.
We deploy \glspl{enodeb} in a hexagonal grid with omnidirectional antennas.
While actual \glspl{enodeb} often use multiple antennas, these antennas are deployed such that they collectively cover a 360\textdegree{} radius around the \gls{enodeb}.
Hence, we can use omnidirectional antennas in our simulation for simplicity without this affecting the outcome.
We repeated each simulation five times and averaged the results.
Each simulation lasts 5\,s, which suffices for all \glspl{ue} to complete the random access procedure in all scenarios.
We tested our simulations with frequency bands 17 (center frequency 740\,Mhz), 11 (center frequency 1486\,Mhz), and 7\,(center frequency 2655\,Mhz),
and the outcomes were similar.
Thus, we present results only for frequency band 17.
We define two \gls{prach} configuration schemes: \textit{same} and \textit{different}.
We have simulated with multiple configurations; however, for the scope of this paper, we present the same and different configuration results to illustrate the impact of configuration choices on system performance. The remaining configurations produced intermediate results that fall between these extremes.
In \textit{same}, all cells use the same \textit{PRACH-ConfigIndex}%
\footnote{We used \textit{PRACH-ConfigIndex} value 1 for the \textit{same} configuration scheme because this value occurred most often in our measurements.}.
In \textit{different}, neighboring cells always use a different \textit{PRACH-ConfigIndex}.
In all simulations, we distribute the \glspl{ue} uniformly across the simulated area, and all \glspl{ue} try to connect to a cell at the same time%
\footnote{More \gls{ue}s could be already connected and transmit data; we simulate only \gls{ue}s that want to initiate a new connection.}.

\subsection{Simulation Results}
\begin{figure}
    \centering
    \includegraphics[width=\columnwidth]{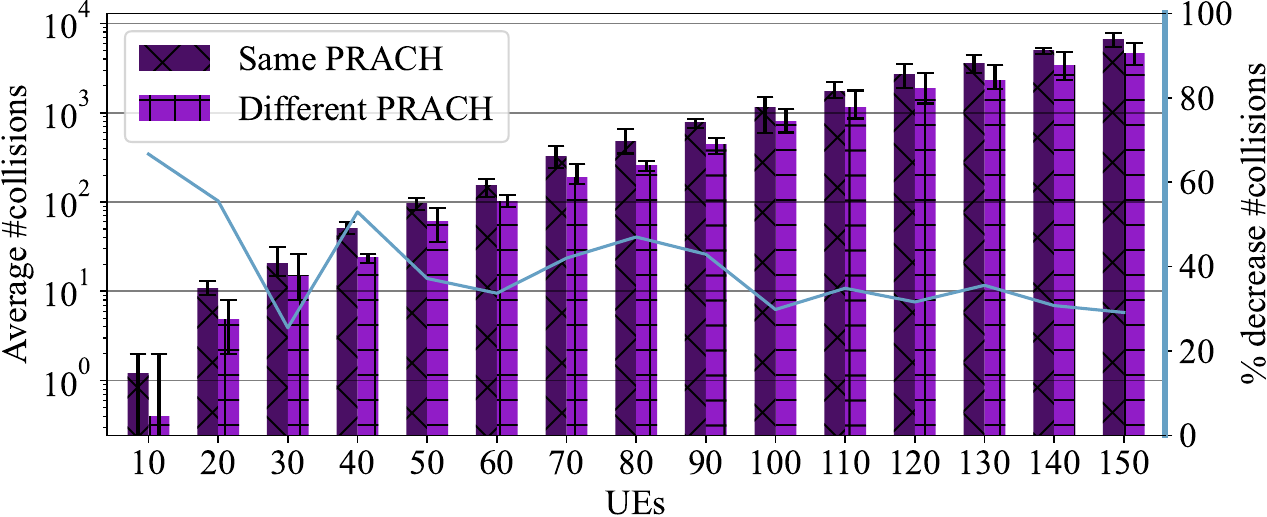}
    \caption{The average number of collisions per amount of \gls{ue}s on a log-scale axis (left y-axis) and the percentual decrease when using different configurations (right y-axis) in the \textit{2-cell scenario}.
    In a simple scenario, using different configurations lowers the number of collisions significantly.}
    \label{fig:avg-collisions-2-cell}
\end{figure}

\noindent\textbf{2-cell Scenario.}\quad
We first simulate a \textit{2-cell scenario} to see how much the network performance is affected by the choice of \gls{prach} configuration in the most straightforward scenario in which contention is possible.
We vary the number of \glspl{ue} that try to connect to a cell at the same time to compare different amounts of random access messages.
Few connecting \glspl{ue} could represent regular traffic where some people move into a different cell, and many connecting \glspl{ue} could represent an arriving train or machine-type devices such as smart meters trying to join the network.
\Cref{fig:avg-collisions-2-cell} shows for varying numbers of \glspl{ue} the average number of collisions across five repetitions of the simulation, with error bars indicating the entire range of observed values.
It compares the results using the \textit{same} and \textit{different} \gls{prach} configuration schemes and the corresponding percentual decrease in the average number of collisions between the two configurations.
Simply changing the configuration from \textit{same} to \textit{different}
reduces the average number of collisions by 40\% on average, and we observed reductions of up to 67\%.
While the percentage by which collisions decrease is lower when there are more \glspl{ue} in total, the \emph{absolute} decrease in collisions is larger with more \glspl{ue}.
Hence, the impact of using different configurations is greater the more \glspl{ue} try to connect at the same time.

\begin{figure}
    \centering
    \includegraphics[width=\columnwidth]{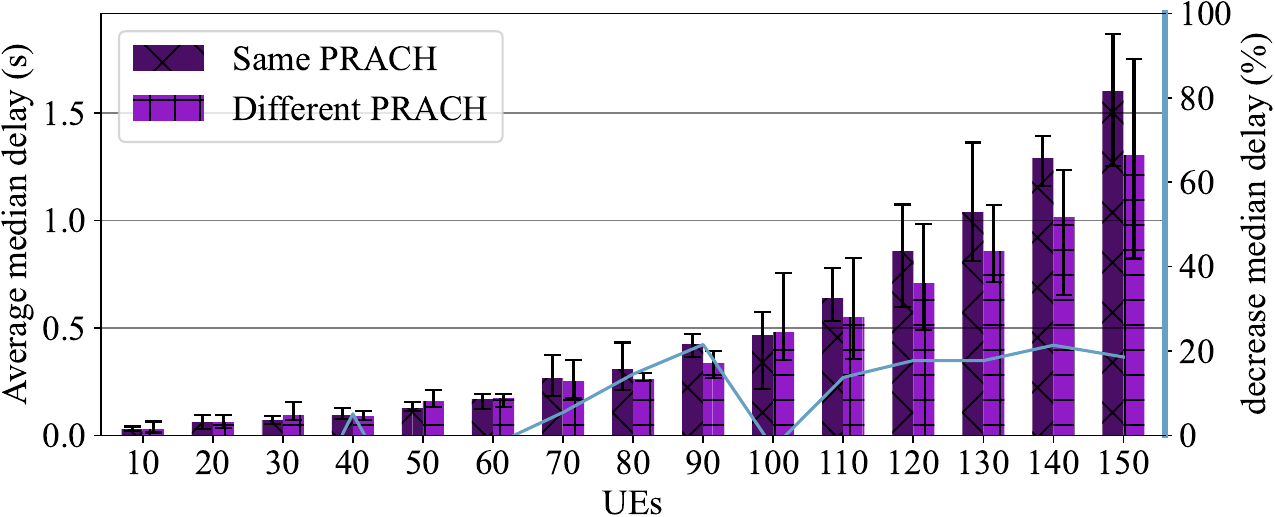}
    \caption{The average median connection delay (median across \glspl{ue}, average across simulations) per amount of \glspl{ue} (left y-axis) and the percentual decrease when using different configurations (right y-axis) in the \textit{2-cell scenario}.
    In a simple scenario, using different configurations reduces the connection delay notably.}
    \label{fig:avg-median-delay-2-cell}
\end{figure}

Similarly, \Cref{fig:avg-median-delay-2-cell} illustrates for varying numbers of \glspl{ue} and for the \textit{same} and \textit{different} \gls{prach} configurations the median connection delay of all \glspl{ue} within one simulation, averaged over five repetitions of the simulation.
It further presents the corresponding percentual decrease in the average median delay between the two configurations.
Using different \textit{PRACH-ConfigIndex} values results in a substantial decrease in the connection delay for the \glspl{ue}.
On average, the connection delay decreases by 5\% and this decrease tends to be larger when there are more \glspl{ue}.
In the worst-case scenario, when the total number of \glspl{ue} is largest, the median connection delay decreases by 300\,ms or 19\%.

\noindent\textbf{19-cell Scenario.}\quad
The \textit{2-cell scenario} provides interesting insights, but actual deployments comprise many cells not just two.
Therefore, we simulate a \textit{19-cell scenario}, which represents a larger area where \glspl{ue} try to connect at the same time.
We chose 19 cells because 3GPP defines 19 as a suitable number of cells for simulations of cellular networks~\cite{tr125-942}.
As before, we evaluate the number of collisions and the connection delay for varying numbers of \glspl{ue} and compare the results when using the \textit{same} or \textit{different}%
\footnote{To ensure that neighboring cells in a large hexagonal grid are always assigned different \textit{PRACH-ConfigIndex} values out of the set of four available values, we use the following assignment scheme.
We split the hexagonal grid into rows (horizontally neighboring cells).
On each row, we alternate between two values, using two distinct sets of values for alternating rows.
For instance, we assigned 0 and 1 to even rows and 2 and 15 to odd rows.
When reusing a set of values, we invert the order.
In our case, we assign row 0 the values 0, 1, 0, 1, \ldots, row 2 the values 1, 0, 1, 0, \ldots, and so forth.}
configurations.

\begin{figure}
    \centering
    \includegraphics[width=\columnwidth]{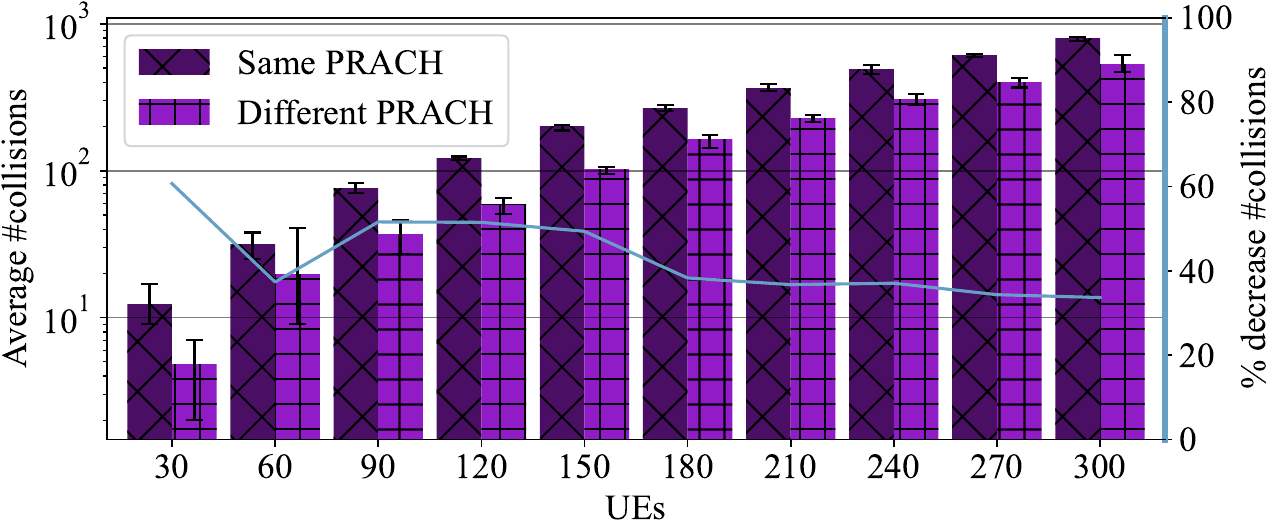}
    \caption{The average number of collisions per amount of \glspl{ue} on a log-scale axis (left y-axis) and the percentual decrease when using different configurations (right y-axis) in the \textit{19-cell scenario}.
    In a large area, assigning different configurations to neighboring cells lowers the number of collisions significantly.}
    \label{fig:avg-collisions-19-cell}
\end{figure}

\Cref{fig:avg-collisions-19-cell} shows the average number of collisions in our \textit{19-cell scenario}.
We obtain similar results as in the \textit{2-cell scenario}: the average reduction in the number of collisions when using different \textit{PRACH-ConfigIndex} values compared to using the same value is 43\%,
and we observed reductions of up to 61\%.
Again, the relative decrease in collisions is smaller when there are more \glspl{ue}, but the absolute decrease is larger.

\begin{figure}
    \centering
    \includegraphics[width=\columnwidth]{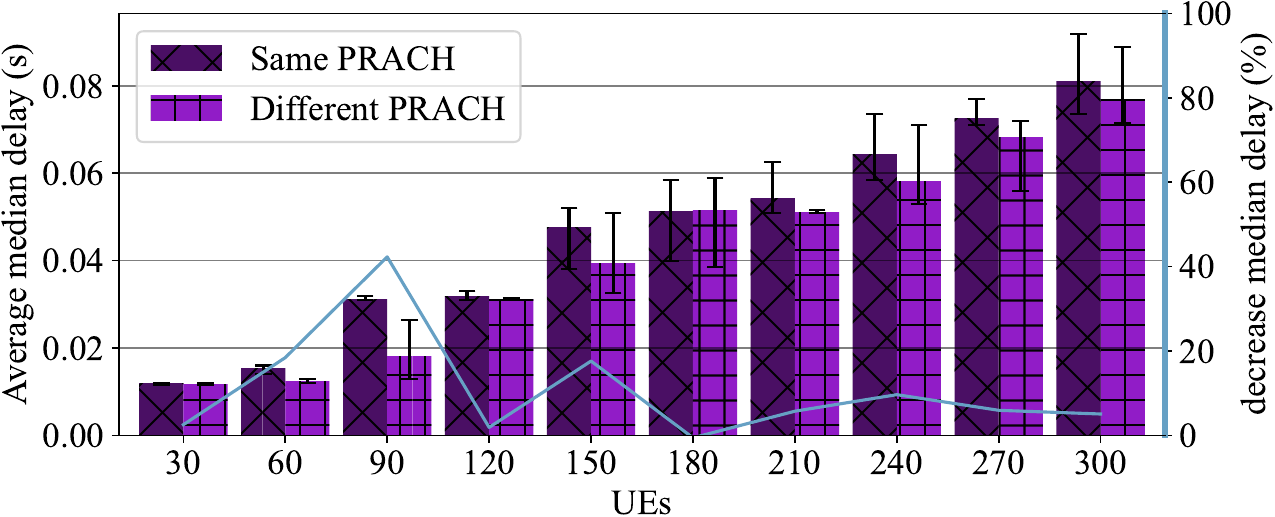}
    \caption{The average median connection delay (median across \glspl{ue}, average across simulations) per amount of \glspl{ue} (left y-axis) and the percentual decrease when using different configurations (right y-axis) in the \textit{19-cell scenario}.
    In a large area, assigning different configurations to neighboring cells reduces the connection delay notably.}
    \label{fig:avg-median-delay-19-cell}
\end{figure}

\Cref{fig:avg-median-delay-19-cell} illustrates the median connection delay of all \glspl{ue} within one simulation, averaged over five repetitions of the simulation.
On average, using different \textit{PRACH-ConfigIndex} values results in an 11\% decrease in the connection delay, and the decrease reaches up to 42\%.
Our findings suggest that in sizeable deployments of many cells,
using different \textit{PRACH-ConfigIndex} values yields substantial performance benefits for any number of \glspl{ue}.
While our simulations cannot represent all details of real-life deployments, they do show that there is potential for improving the performance of cellular networks by simply deploying different configurations.

%% file: sections/relatedWork.tex
\section{Related Work}
\label{sec:related-work}

\noindent\textbf{Measurements.}\quad
Numerous studies have conducted real-life measurements in cellular networks from the point of view of a \textit{\gls{ue}}.
They offer insights into
mobility patterns~\cite{kolamunna2018first, andrade2017connected},
power consumption~\cite{yang2020understanding, xu2020understanding},
network coverage and performance~\cite{ghoshal2023performance},
handover performance~\cite{andrade2017connected, xu2020understanding, li2018measurement, deng2018mobility},
end-to-end latency~\cite{xu2020understanding},
software reliability for connection failures~\cite{li2021nationwide},
and overall \gls{qoe}~\cite{xu2020understanding}.
Due to the absence of proper measurement tools from \gls{mno}s, researchers have created \textit{measurement platforms}~\cite{nikravesh2015mobilyzer, li2016mobileinsight, li2021experience, alay2017experience, bonati2021scope} to better enable the community to conduct measurements in cellular networks.
Researchers have also developed applications for analyzing low-level protocol operations in 4G that run on top of \gls{ue}s~\cite{peng2016understanding, shi2022towards}.
Furthermore, there are \textit{measurement studies by \gls{mno}s} that have delved into
roaming with smartphones~\cite{mandalari2018experience} or \gls{iot} devices~\cite{lutu2020things}, handovers and user mobility~\cite{lutu2020characterization},
and diagnosing \gls{ran} problems~\cite{dasari2018impact, iyer2017automating}.
Unfortunately, \gls{mno}s rarely make data openly available. 
For random access configurations in particular, there exist, to the best of our knowledge,
no previous studies that have collected, analyzed, and published real-life data.

\noindent\textbf{\gls{ran} Configurations.}\quad
Researchers have analyzed the impact of \textit{misconfigurations} that lead to handover loops, both within one and between different radio access technologies~\cite{zhang2023dependent, zhao2018measuring, li2016instability} as well as between small cells and macro-cells~\cite{peng2016understanding}.
Such loops cause a \gls{ue} to keep performing handovers even if the radio conditions or the \gls{ue}'s location do not change.
Misconfigurations can also lead to situations where a \gls{ue} performs a handover, but the newer cell performs worse than the previous cell~\cite{zhang2023dependent}.
With regard to random access procedures, we are the first to inspect how cellular networks configure these procedures and find how real-world configurations are suboptimal.

Moreover, researchers have made use of real-life 4G and 5G datasets for \textit{configuration recommendation systems} such as Aurora~\cite{mahimkar2022aurora} and SOCRATES~\cite{eisenblatter2011self}.
These systems use information on the configuration and current performance to recommend or automatically tune network parameters in an attempt to improve the network performance.
Such configuration recommendation systems can make use of insights from the random access configurations, and could leverage measurements such as ours to strengthen their recommendations and, thus, improve the overall network performance.

\noindent\textbf{\gls{rrc} Procedures.}\quad
Some research has aimed to optimize \gls{rrc} procedures~\cite{mahmood2016radio, stoner2015experience, wei2016measuring, li2016icellular} including random access procedures~\cite{lenaplus, huang2017r}.
Other work has focused on optimizing \gls{rrc} algorithms for handovers~\cite{liu2008performance, brunner2006inter, lobinger2011coordinating},
improving energy usage~\cite{koc2014device},
and cell selection strategies~\cite{coronado2022roadrunner}.
Nevertheless, none of these works make use of real-life datasets to show the potential impact of configurations in actual cellular deployments.
We believe that such insights are vital for further improving the design of \gls{rrc} schemes.